\documentclass[prl,superscriptaddress,preprint,endfloats]{revtex4}

\newcommand {\CA}{Cd$_3$As$_2${}}
\newcommand {\FS}{Fe$_3$Sn$_2${}}
\newcommand {\ECS}{EuCd$_2$Sb$_2${}}
\newcommand {\EZS}{EuZn$_2$Sb$_2${}}
\newcommand {\SRO}{SrRuO$_3${}}

\newcommand {\sigmaxx}{$\sigma _{xx}${}}

\newcommand {\sigmaxyAHE}{$\sigma _{xy,{\mathrm{AHE}}}${}}

\newcommand {\rhoxx}{$\rho _{xx}${}}
\newcommand {\rhoyx}{$\rho _{yx}${}}
\newcommand {\rhoyxsym}{$\rho _{yx,\mathrm{sym}}${}}
\newcommand {\rhoyxasym}{$\rho _{yx,\mathrm{asym}}${}}
\newcommand {\rhoyxAHE}{$\rho _{yx,{\mathrm{AHE}}}$}

\newcommand{\bm}[1]{{\mbox{\boldmath $#1$}}}

\usepackage{graphicx}
\usepackage{bm} 
\usepackage{pifont}
\usepackage{amsmath, amsthm, amssymb}
\usepackage{color}
\usepackage{dcolumn} 

\begin{document}
\title{Anomalous Hall effect in Dirac semimetal probed by in-plane magnetic field}
\author{Shinichi Nishihaya}
\affiliation{Department of Physics, Institute of Science Tokyo, Tokyo 152-8551, Japan}
\author{Hiroaki Ishizuka}
\affiliation{Department of Physics, Institute of Science Tokyo, Tokyo 152-8551, Japan}
\author{Yuki Deguchi}
\affiliation{Department of Physics, Institute of Science Tokyo, Tokyo 152-8551, Japan}
\author{Ayano Nakamura}
\affiliation{Department of Physics, Institute of Science Tokyo, Tokyo 152-8551, Japan}
\author{Tadashi Yoneda}
\affiliation{Department of Physics, Institute of Science Tokyo, Tokyo 152-8551, Japan}
\author{Hsiang Lee}
\affiliation{Department of Physics, Institute of Science Tokyo, Tokyo 152-8551, Japan}
\author{Markus Kriener}
\affiliation{RIKEN Center for Emergent Matter Science (CEMS), Wako 351-0198, Japan}
\author{Masaki Uchida}
\email[Author to whom correspondence should be addressed: ]{m.uchida@phys.sci.isct.ac.jp}
\affiliation{Department of Physics, Institute of Science Tokyo, Tokyo 152-8551, Japan}

\begin{abstract}
\textbf{Intrinsic anomalous Hall effect (AHE) formulated by geometric properties of Bloch wavefunctions is a ubiquitous transport phenomenon not limited to magnetic systems but also allowed in non-magnetic ones under an external field breaking time-reversal symmetry. On the other hand, detection of field-induced AHE is practically challenging because the band modulation through the Zeeman and spin-orbit couplings is typically small compared to other contributions as induced by the Lorentz force. Here, we demonstrate on Dirac semimetal {\CA} films that the field-induced AHE in non-magnetic systems can be quantitatively probed by applying and rotating the magnetic field within the Hall deflection plane. Measurements on the {\CA} (112) plane reveal that AHE emerges as a clear three-fold symmetric component for the in-plane field rotation. This intrinsic response becomes more pronounced in ultralow-electron-density films where significant variations in the geometric properties are expected under the magnetic field. Our findings open new opportunities in the research of Hall responses manifested as orbital magnetization in non-magnetic systems.}
\end{abstract}
\maketitle

In recent condensed matter physics, anomalous Hall effect (AHE) has been under intense focus as a measure of the geometric properties of the Bloch wavefunctions. Contrary to the early phenomenological belief that AHE is proportional to the out-of-plane spin magnetization, the modern formulation based on Berry curvature \cite{AHEreview} has clarified that it is not limited to systems with sizable magnetization, but also allowed in antiferromagnets with a vanishingly small magnetization \cite{AHEAFMreview} and even in non-magnetic systems \cite{BCreview,iAHE_NM1,iAHE_NM2}, as long as the Berry curvature integrated over the occupied states remains finite under time-reversal symmetry broken conditions. In contrast to AHE in magnetic systems with spontaneous magnetization (Fig. 1\textit{A}), emergence of AHE in non-magnetic systems should rely on the generation of Berry-curvature related orbital magnetization by external fields \cite{orbitalmag_rev,orbitalmag1,orbitalmag2}. One related example is the non-linear Hall effect reported in inversion-symmetry-broken systems \cite{NHE1,NHE2}. The non-zero Berry curvature dipole combined with an electric field applied normal to it induces an out-of-plane orbital magnetization and Hall response under zero field \cite{NHE1,NHE2}. On the other hand, orbital magnetization as a direct consequence of non-zero Berry curvature can also be generated by a magnetic field through Zeeman and spin-orbit couplings as shown in Fig. 1\textit{B}. In principle, the magnetic-field-induced AHE should appear in non-magnetic systems, which however has rarely been reported due to its small amplitude typically overwhelmed by other contributions as induced by the Lorentz force. 

As for the experimental detection of the magnetic-field-induced AHE, a sufficiently large modulation of the band structure by the magnetic field is required. In this respect, promising candidates are gapless semimetals or narrow-gapped semiconductors with large $g$-factor which can form topological band (anti-)crossings hosting giant Berry curvature under the magnetic field. In fact, unconventional responses attributed to the anomalous Hall or Nernst effect have been reported in pioneering studies of non-magnetic systems such as the Dirac semimetal {\CA} \cite{ANE_CA} and the massive Dirac semimetal ZrTe$_5$ \cite{AHE_ZT1,AHE_ZT2}. However, the observed field dependence implies a coexistence of different contributions, hindering the quantification of AHE. Actually, it is challenging to identify the field-induced AHE measured under the conventional out-of-plane field configuration, since the measured transverse response is dominated by Lorentz-force effects. Namely, the field-linear component of AHE is not distinguishable from the single-carrier ordinary Hall effect (OHE). Even if there is a non-linear or nonmonotonic component of AHE, it is practically difficult to exclude possible multi-carrier OHE or other inhomogeneity-induced effects, especially in low-carrier density samples. 

Here we focus on the possibility that the field-induced AHE in non-magnetic systems can be probed by an in-plane magnetic field. With fulfilling symmetry conditions \cite{iAHE_sym1,iAHE_sym2}, out-of-plane Berry curvature and associated orbital magnetization can be induced even by the in-plane field, leading to in-plane AHE as illustrated in Fig. 1\textit{C}. Being a field-odd effect, in-plane AHE is only allowed on the crystal plane with odd-fold rotational symmetry about its out-of-plane axis \cite{iAHE_sym1,iAHE_sym2}. Following many theoretical proposals \cite{iAHE_NM1,iAHE_NM2,iAHE_NM3,iAHE_the1_Q,iAHE_the2_Q,iAHE_the3_Q,iAHE_the4_Q,iAHE_the5,iAHE_the6,iAHE_the7}, in-plane AHE with three-fold rotational symmetry has been reported in various magnetic systems \cite{iAHE_ECS, iAHE_EZS, iAHE_Fe, iAHE_FeSn, iAHE_SRO}. The field-induced Hall response up to the $B^1$ order with one-fold rotational symmetry has been observed in several non-magnetic systems \cite{AHE_ZT1,iAHE_VS,iAHE_STO}, and it has been proposed to link to the quantum metric quantities of the band structure as formulated by extended semiclassical theories \cite{iAHE_NM2,iAHE_the6,iAHE_the7}. On the other hand, the in-plane AHE proportional to $B^{3}$ on a plane with $C_3$ symmetry, such as the (001) plane of a trigonal system \cite{iAHE_ECS, iAHE_EZS,iAHE_FeSn} or the (111) plane of a cubic system \cite{iAHE_SRO} offers exceptional advantages for quantifying the in-plane AHE. The anomalous Hall response with a three-fold symmetric feature for the in-plane field rotation can be easily separated from the Lorentz-force induced OHE either due to the slight sample misalignment or due to the selection of a low-symmetry plane with non-zero off-diagonal components in the Hall conductivity tensor \cite{iOHE1,iOHE2}, all of which contribute to the one-fold symmetric component. 

In this study, we report observation of in-plane AHE in Dirac semimetal {\CA} films. We have performed measurements on the (112) plane of {\CA} with the tetragonal $D_{4h}$ structure \cite{CA_crystal}, which possibly allows the one-fold and three-fold components of in-plane AHE. Figures 1\textit{D}-1\textit{F} show the field-induced Hall conductivity on the (112) plane calculated using an effective model of {\CA} constructed from previous first-principles calculations \cite{Wang2013a,Arribi2020a,Baidya2020a,Smith2024a} (see Materials and Methods for details). Notably, a field applied along the in-plane direction [$11\bar{4}$] or [$\bar{2}12$] is expected to induce AHE with an amplitude comparable to that induced by the out-of-plane field $B\parallel[112]$. The three-fold component of in-plane AHE corresponds to the average of the three cases with the in-plane field rotated by $120^{\circ}$ between each other (dashed line in Fig. 1\textit{F}), which is about one order of magnitude smaller but still remains finite. Through measurements on low-electron-density {\CA} films, we have succeeded in identifying in-plane AHE appearing with a three-fold component following the crystal symmetry. Reflecting its intrinsic origin, in-plane AHE is more pronounced with larger Hall angles in samples with electron density below $10^{17}$ cm$^{-3}$. Our demonstration of in-plane AHE broadens the research of Hall physics in non-magnetic systems, potentially leading to new frontiers focusing on in-plane field-induced orbital magnetization. 

\section*{Results}
\subsection*{Fundamental structural and transport properties}
(112) {\CA} films were grown on (111)A CdTe substrate by molecular beam epitaxy (see Materials and Methods for details). Figure 2 summarizes structural and transport properties of the {\CA} films. For the epitaxial stacking of (112) {\CA} on (111)A CdTe, there are three equivalent configurations with either {\CA}\ $[11\bar{4}]$, $[\bar{2}12]$, or  $[1\bar{2}2]$ parallel to CdTe $[11\bar{2}]$ (Fig. 2\textit{A}). While it is difficult to identify such $120^{\circ}$-rotated domains by standard x-ray diffraction (XRD) due to the nearly identical lengths of the $c$-axis and the doubled $a$- and $b$-axes, their presence has been reported in the previous study using the convergent beam electron diffraction technique \cite{CA_domains}. Throughout this work, we do not distinguish these three in-plane domains since they do mutually attenuate the one-fold component of in-plane AHE but contribute equally to the three-fold one. As presented in Fig. 2\textit{B}, the XRD $\varphi$-scan of {\CA} $(664)$ (or $(62\underline{12})$, $(26\underline{12})$) Bragg peak shows that the formation of $60^{\circ}$-rotated twins is highly suppressed in the present {\CA} films. This is crucial for observing the three-fold component of in-plane AHE.  

Figures 2\textit{C} presents temperature dependence of resistivity {\rhoxx} for the {\CA} films with different thicknesses and electron densities (see \textit{SI Appendix}, Table S1 for the details of the samples). Each sample exhibits a different electron density ranging from $4.6\times10^{16}$ to $7.9\times10^{17}$ cm$^{-3}$. These low-electron-density {\CA} films typically exhibit a semiconducting temperature dependence of {\rhoxx}, consistent with the previous report \cite{CA_low}. Figures 2\textit{D} and 2\textit{E} present Hall resistivity {\rhoyx} and magnetoresistivity ratio $\rho_{xx}(B)/\rho_{xx}(0)$ data measured under the out-of-plane field. While thicker films (Samples A, C, and E) show positive magnetoresistance accompanied by quantum oscillations, thinner films (Samples B and D) exhibit clear development of confinement-induced quantum Hall states \cite{CA_2D} with the filling factor $\nu = 1$ or $2$ at the high fields. As discussed later, the observed amplitude of in-plane AHE is sensitive to the electron density while the confinement-induced dimensionality change has only a negligible effect.  

\subsection{In-plane anomalous Hall effect}
Next we present the Hall responses measured under the in-plane field for Samples A and B. The azimuthal angle $\varphi$ of the in-plane field is measured from the {\CA} $[11\bar{4}]$ direction. First we discuss the results of sample A measured by applying the current along the $[1\bar{1}0]$ direction ($\varphi=270^{\circ}$) as illustrated in Fig. 3\textit{A}. Generally, {\rhoyx} measured under the in-plane field involves a magnetoresistance effect, namely, planar Hall effect  in its field-symmetric part {\rhoyxsym}= [{\rhoyx}($B$)+{\rhoyx}($-B$)]/2 = [{\rhoyx}($\varphi$)+{\rhoyx}($\varphi+180^{\circ}$)]/2, and OHE and AHE in its field-asymmetric part {\rhoyxasym} = [{\rhoyx}($B$)$-${\rhoyx}($-B$)]/2 = [{\rhoyx}($\varphi$)$-${\rhoyx}($\varphi+180^{\circ}$)]/2. To further extract the three-fold component {\rhoyxAHE}, we have subtracted the one-fold component from {\rhoyxasym}, which is mainly derived from the misalignment-induced out-of-plane OHE (see \textit{SI Appendix} for the detailed procedure). 

As clearly seen in Fig. 3\textit{B}, the three-fold component emerges as the leading term of in-plane AHE. We note that the three-fold $\varphi$ dependence can be captured even before subtracting the one-fold component (\textit{SI Appendix}, Figs. S2 and S4). The {\rhoyxAHE} curves measured at high fields exhibit positive and negative extrema at $\varphi = 0^{\circ}$ and $60^{\circ}$ and their equivalent directions. Since we focus on the three-fold component, the observed $\varphi$ dependence can be understood based on symmetry considerations of the high-temperature {\CA} phase with the cubic anti-fluorite structure (point group $Fm\bar{3}m$)  \cite{CA_crystal2}. Consistent with the symmetry requirements \cite{iAHE_sym1,iAHE_sym2}, in-plane AHE is prohibited when the field is applied to the $[01\bar{1}]_{\mathrm{c}}$ ($\varphi = 30^{\circ}$) or its equivalent directions. In each of these directions, there are a $C_2$ rotational symmetry axis along it and a mirror symmetry plane perpendicular to it. 

In order to confirm the reproducibility, we have measured Sample B with applying the current to a different direction ($\varphi = 0^{\circ}$) rotated by $90^{\circ}$ as compared to the case of Sample A, as shown in Fig. 3\textit{D}. The {\rhoyxAHE} curves obtained for Sample B in Fig. 3\textit{E} exhibit the same $\varphi$ dependence as found for Sample A. This highlights the key feature of in-plane AHE being strictly constrained by the crystal symmetry, and hence can be easily distinguished from other one-fold effects.

\subsection*{Magnetic field and electron density dependences}

Magnetic field dependence of {\rhoyxAHE} is also obtained by subtracting the one-fold component from the {\rhoyxasym} curve with referring to the $\varphi$ scans (see \textit{SI Appendix} for the detailed procedure). As presented in Figs. 3\textit{C} and 3\textit{F}, {\rhoyxAHE} of both Samples A and B exhibits a similar unique field dependence, which is composed of an oscillatory behavior below about 5 T and a large enhancement above. This field dependence can be also confirmed in the $\varphi$ scans in Figs. 3\textit{B} and 3\textit{E}.

Since the three-fold component effectively emerges as an average of field-induced higher-order effects in the $120^{\circ}$-rotated domains, its field dependence could be nonmonotonic depending on details of the field-induced band modulation and the Fermi level position. Our model calculation of the three-fold component in Fig. 1\textit{F} also shows nonmonotonic field dependence with a sign change, while only a monotonic behavior is expected for each domain. This nonmonotonic in-plane field dependence seems not simply related to the Landau level splitting, considering that it appears in Samples A and B regardless of the dimensionality of the electronic structure. Thus, its detailed interpretation remains for future studies.    

Figure 4\textit{A} presents temperature dependence of {\rhoyxAHE} curves for Sample A. The amplitude decays monotonically with increasing temperature, indicating that the field-induced Berry curvature effect is suppressed by increased scattering and thermal broadening of the bands at higher temperatures. Figure 4\textit{B} compares the {\rhoyxAHE} curves taken for {\CA} films with different electron densities at $\varphi = 180^{\circ}$ or its equivalent directions. As represented by Sample C, films with electron density higher than $1\times10^{17}$ cm$^{-3}$ exhibit only a small amplitude originating from in-plane AHE (see also \textit{SI Appendix}, Fig. S4). The in-plane anomalous Hall angle estimated from $\sigma_{xy,\mathrm{AHE}}/\sigma_{xx}$ also exhibits the overall trend that it is more enhanced upon decreasing the electron density (\textit{SI Appendix}, Fig. S4). This highlights the intrinsic origin of the observed in-plane AHE. Namely, it is expected to be enhanced as the Fermi level approaches the band degenerate points or band bottoms which become Berry curvature hot spots under the magnetic field. 

\section*{Discussion}

Figure 4\textit{C} shows the scaling relation between the anomalous Hall conductivity {\sigmaxyAHE} and the longitudinal conductivity {\sigmaxx} obtained for the {\CA} films. The in-plane anomalous Hall angle of the ultralow-electron-density films (Samples A and B) is even larger as compared to previously reported magnetic materials \cite{iAHE_ECS,iAHE_EZS,iAHE_FeSn,iAHE_Fe,iAHE_SRO}, and it reaches 2.4\% at 2 K. Note that the in-plane AHE of {\CA} quantified here is only the three-fold component. If a single crystalline sample without $120^{\circ}$-rotated domains could be prepared, even larger in-plane AHE with one-fold component can be expected as calculated in Fig. 1\textit{F}. Our findings provide a new route for exploring the intrinsic Hall response even in non-magnetic systems. Quantification of in-plane AHE largely dependent on the carrier density also potentially leads to the demonstration of its quantization as theoretically proposed in previous studies \cite{iAHE_the1_Q,iAHE_the2_Q,iAHE_the3_Q,iAHE_the4_Q,iAHE_NM3}.  

In summary, we have extracted large AHE in Dirac semimetal {\CA} films through systematic measurements under the in-plane magnetic fields. The emergence of the in-plane AHE strictly follows the crystal symmetry of the {\CA} (112) plane, regardless of the current direction. The obtained in-plane anomalous Hall angle exceeds that of magnetic systems reported previously, suggesting the importance of tuning the Fermi level to achieve large field-induced Berry curvature. The in-plane AHE in non-magnetic systems can be interpreted as the effect of field-induced orbital magnetization. Together with recent research advances concerning the non-linear Hall effect \cite{NHE1,NHE2}, this work demonstrates that the utilization of orbital magnetization leads to exploration of novel physics and functionalities even in non-magnetic systems. 

\section*{Methods and Materials}
\subsection*{Sample preparation}
[112]-oriented {\CA} films were grown by molecular beam epitaxy (MBE) on CdTe ($111$)A surface in an EpiQuest RC1100 chamber. Elemental sources were co-evaporated using a conventional Knudsen cell for Cd (6N, Osaka Asahi Co.) and an MBE-Komponenten valved cracker source for As (7N5, Furukawa Co.). The reservoir temperature of the cracker source was set to $950^{\circ}$C to sublimate arsenic in the form of As$_2$. Prior to the growth of {\CA}, CdTe substrates were etched with 0.01\% Br$_2$-methanol for 5 min to remove the native oxide layer, and then annealed at $520$-$560^{\circ}$C under Cd overpressure inside the MBE chamber. The film thicknesses were designed to be $10$-$100$ nm. Samples with different electron densities were prepared by adjusting the growth temperature ($190$-$210^{\circ}$C) and the Cd/As$_2$ flux ratio ($10$-$40$). 
\subsection*{Transport measurement}
Transport measurements in the temperature range from 300 K to 2 K and the field range up to 9 T were performed in a Quantum Design Physical Property Measurement System (PPMS) or Cryomagnetics cryostat system equipped with a superconducting magnet. The field-angle dependent measurements were obtained by using a sample rotator. 4-probe resistance was measured on Hall-bar devices using standard DC or low-frequency lock-in techniques.
\subsection*{Model calculation of anomalous Hall conductivity}
For the calculation of the anomalous Hall conductivity of \CA, we used an effective model for Cd$_3$As$_2$ constructed from the first-principles band calculations~\cite{Wang2013a,Arribi2020a,Baidya2020a,Smith2024a}. The Hamiltonian reads
\begin{align*}
H({\bm k})=&H_0({\bm k})+H_h({\bm k}),\\
H_{0}({\bm k})=&\epsilon_{0}({\bm k})\tau_{0}s_{0}+\mathcal{M}({\bm k})\tau_{z}s_{0}+\,A\,(k_{x}\tau_{x}s_{z}-k_{y}\tau_{y}s_{0})+\,C_{3a}k_{x}k_{y}\,(k_{x}\tau_{y}s_{0}-k_{x}\tau_{x}s_{z})\\
&+\,C_{3b}\left(k_{x}^{3}\tau_{x}s_{z}-k_{y}^{3}\tau_{y}s_{0}\right)+\,C_{3c}k_{z}^{2}\left(k_{y}\tau_ys_{0}-k_{x}\tau_{x}s_{y}\right) +\,C_{3d}(k_{x}^{2}-k_{y}^{2})k_{z}\tau_{x}s_{x}\\
&-\,C_{3e}k_xk_yk_z\tau_xs_{y},\\
H_h(\bm k)=&\left(\frac{\tau_{0}+\tau_{z}}{2}\right)\beta_s\bm h\cdot\mathbf{s}+\left(\frac{\tau_{0}-\tau_{z}}{2}\right) \left[\beta_p' h_{x}\left(h_{x}^{2}-3h_{y}^{2}\right)s_{x}-\,\beta_p' h_{y}\left(h_{y}^{2}-3h_{x}^{2}\right)s_{y}+\beta_p h_z s_z\right],
\end{align*}
where $\bm k=(k_x,k_y,k_z)$ is the momentum of electron, $\bm B=(B_x,B_y,B_z)$ is the magnetic field, $\mu_B$ is the Bohr magneton, $\epsilon_{0}(\bm{k})=C_{0}+C_{1}k_{z}^{2}+C_{2}(k_x^{2}+k_y^2)$, $\mathcal{M}(\bm{k})=M_{0}-M_{z}k_{z}^{2}-M_{x y}(k_x^{2}+k_y^2)$, and $\tau_{i}$ ($s_{i}$) are the Pauli matrices acting on the orbital (spin) spaces, respectively.
Following the previous works \cite{Wang2013a,Arribi2020a,Baidya2020a,Smith2024a}, we set the parameters as $C_{0}=-0.0475$ eV, $C_{1}=12.50$ eV\AA$^2$, $C_{2}=13.62$ eV\AA$^2$, $M_{0}=0.0282$ eV, $M_{x y}=13.32$ eV\AA$^2$, $M_{z}=20.72$ eV\AA$^2$, $A=1.116$ eV\AA, $C_{3a}=-600$ eV\AA$^3$, $C_{3b}=-200$ eV\AA$^3$, $C_{3c}=0$ eV\AA$^3$, $C_{3d}=-800$ eV\AA$^3$, $C_{3e}=400$ eV\AA$^3$, $\beta_s=5.4\times10^{-5}$ eVT$^{-1}$, $\beta_p=2\beta_s$, and $\beta_p'=6.15\times10^{-18}$ eVT$^{-3}$. The calculation of the anomalous Hall conductivity is based on the Kubo formula. The integral over the momentum space is taken over $k\le 0.1$ \AA$^{-1}$ using polar coordinates with $N_r=512$, $N_\theta=384$, and $N_\phi=3072$ divisions along the radial, elevation, and azimuth directions, respectively. 

\vskip\baselineskip
\noindent \textbf{Acknowledgement:} 
This work was supported by JST FOREST Program Grant Number JPMJFR202N and PRESTO Program Grant Number JPMJPR2452, by JSPS KAKENHI Grant Numbers JP22K18967, JP22K20353, JP23K13666, JP23K03275, JP24H01614, and JP24H01654 from MEXT, Japan, by Murata Science and Education Foundation, Japan, by STAR Award funded by the Tokyo Tech Fund, Japan, and by Iketani Science and Technology Foundation.
\vskip\baselineskip
\noindent \textbf{Author Contributions:} M.U. conceived the project and designed the experiments. S.N. and Y.D. grew films with A.N., T.Y., and H.L. S.N. and Y.D. performed transport measurements with M.K. H.I. performed the model calculations. S.N. and M.U. wrote the manuscript with input from all authors. All authors have approved the final version of the manuscript.
\vskip\baselineskip
\noindent \textbf{Author Declaration:} The authors declare that they have no competing interests.
\vskip\baselineskip
\noindent \textbf{Key words:} Dirac semimetal, anomalous Hall effect, thin film 
\vskip\baselineskip
\noindent
\textbf{Significance statement:} Anomalous Hall effect (AHE) reflecting geometric properties of electronic wavefunctions has been well studied in magnetic but not in non-magnetic systems. This is because their intrinsic Hall response to the magnetic field is typically disguised by Lorentz force contributions in the conventional out-of-plane field configuration. Notably, however, a comparably sized AHE is theoretically expected even for the in-plane magnetic field. Here, we report quantitative extraction of AHE in non-magnetic Dirac semimetal {\CA} films by focusing on its Hall response to the in-plane field. Measurements with rotating the field within the Hall deflection plane reveal large AHE with a three-fold rotational symmetry following the symmetry requirements. Our findings pave the way for exploring intrinsic Hall responses even in non-magnetic systems.


\begin{figure*}
\begin{center}
\includegraphics[width=16cm]{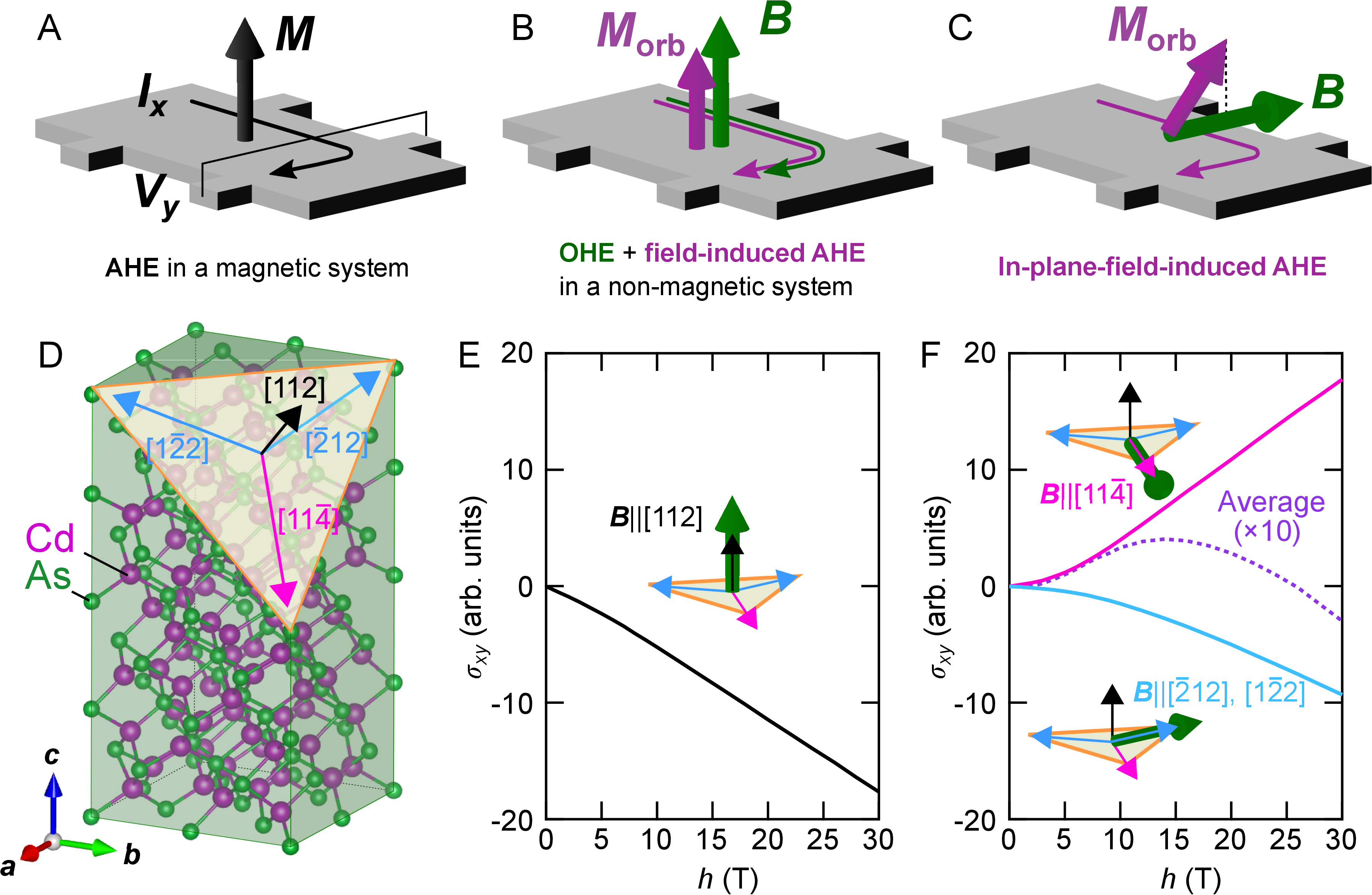}
\caption{Field-induced anomalous Hall effect in non-magnetic {\CA}. (\textit{A}) Anomalous Hall effect (AHE) in a magnetic system with spontaneous magnetization, (\textit{B}) ordinary Hall effect (OHE) and field-induced AHE in a non-magnetic system under the out-of-plane field, and (\textit{C}) AHE induced by an in-plane field applied to the crystal plane with odd-fold rotational symmetry about its out-of-plane direction. (\textit{D}) (112) plane of \CA\ highlighted in its tetragonal structure, which allows the observation of both one-fold and three-fold symmetric in-plane AHE for the in-plane field rotation. Field-induced Hall conductivities calculated for (\textit{E}) the out-of-plane field $B\parallel[112]$ and (\textit{F}) the in-plane fields $B\parallel[\bar{1}\bar{1}4],[2\bar{1}\bar{2}]$ and $[\bar{1}2\bar{2}]$. The dashed line indicates the average of the contributions from the three crystal domains rotated by $120^{\circ}$, which corresponds to the three-fold component of the in-plane AHE. 
}
\label{fig1}
\end{center}
\end{figure*}

\begin{figure*}
\begin{center}
\clearpage
\newpage
\includegraphics[width=16cm]{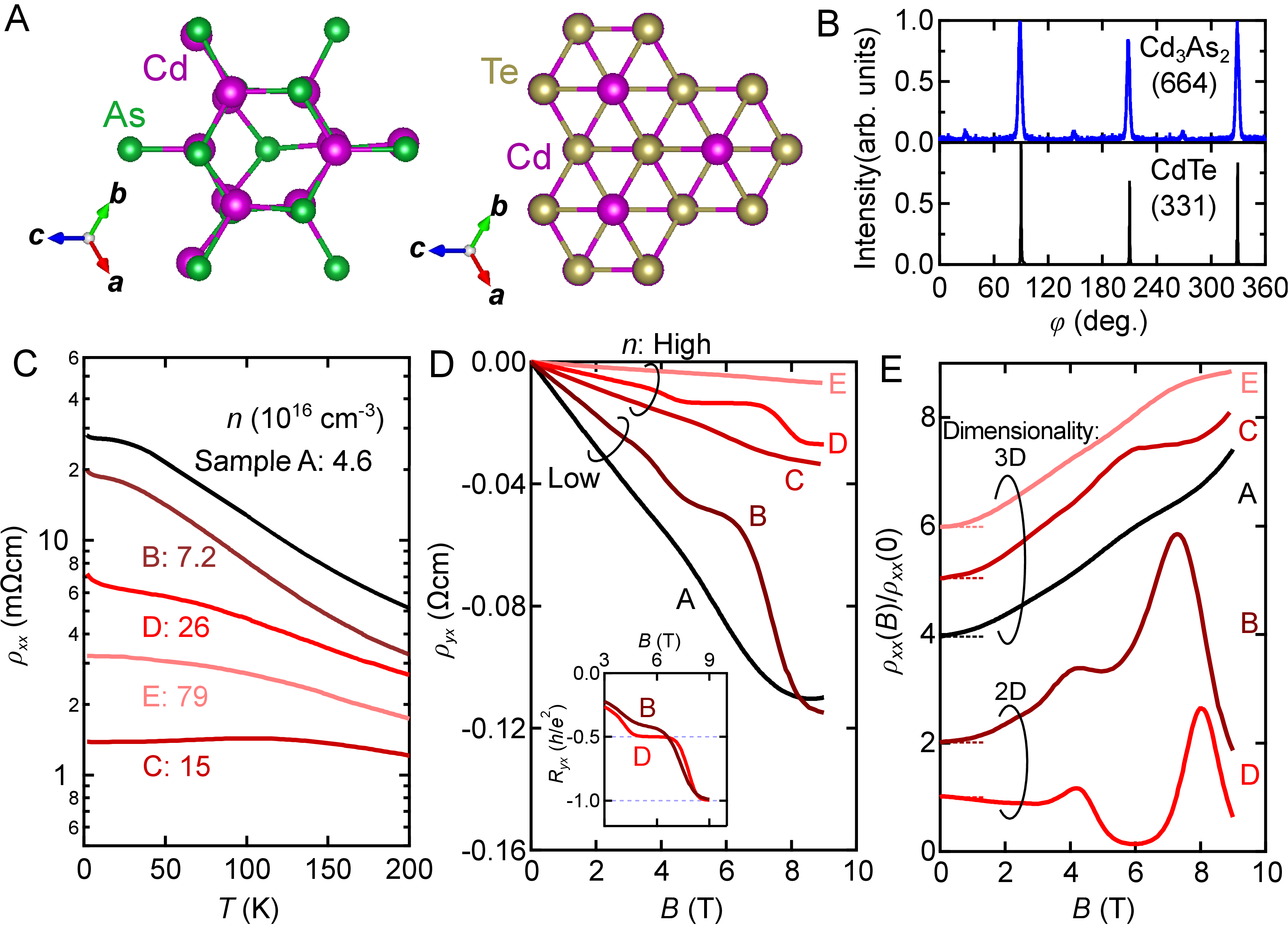}
\caption{Structural and transport characterization of {\CA} films. (\textit{A}) Top view of the {\CA} (112) plane and the CdTe (111)A plane. (\textit{B}) $\varphi$-scan results of the {\CA} (664) and CdTe (331) Bragg peaks. The three-fold pattern indicates the suppression of the formation of 60$^{\circ}$-rotated twins. (\textit{C}) Temperature dependence of longitudinal resistivity {\rhoxx} of {\CA} films with different electron density. Out-of-plane field dependence of (\textit{D}) Hall resistivity {\rhoyx} and (\textit{E}) magnetoresistivity ratio $\rho_{xx}(B)/\rho_{xx}(0)$. Data are shifted for clarity in (\textit{E}).}
\label{fig2}
\end{center}
\end{figure*}

\begin{figure*}
\begin{center}
\clearpage
\newpage
\includegraphics[width=16cm]{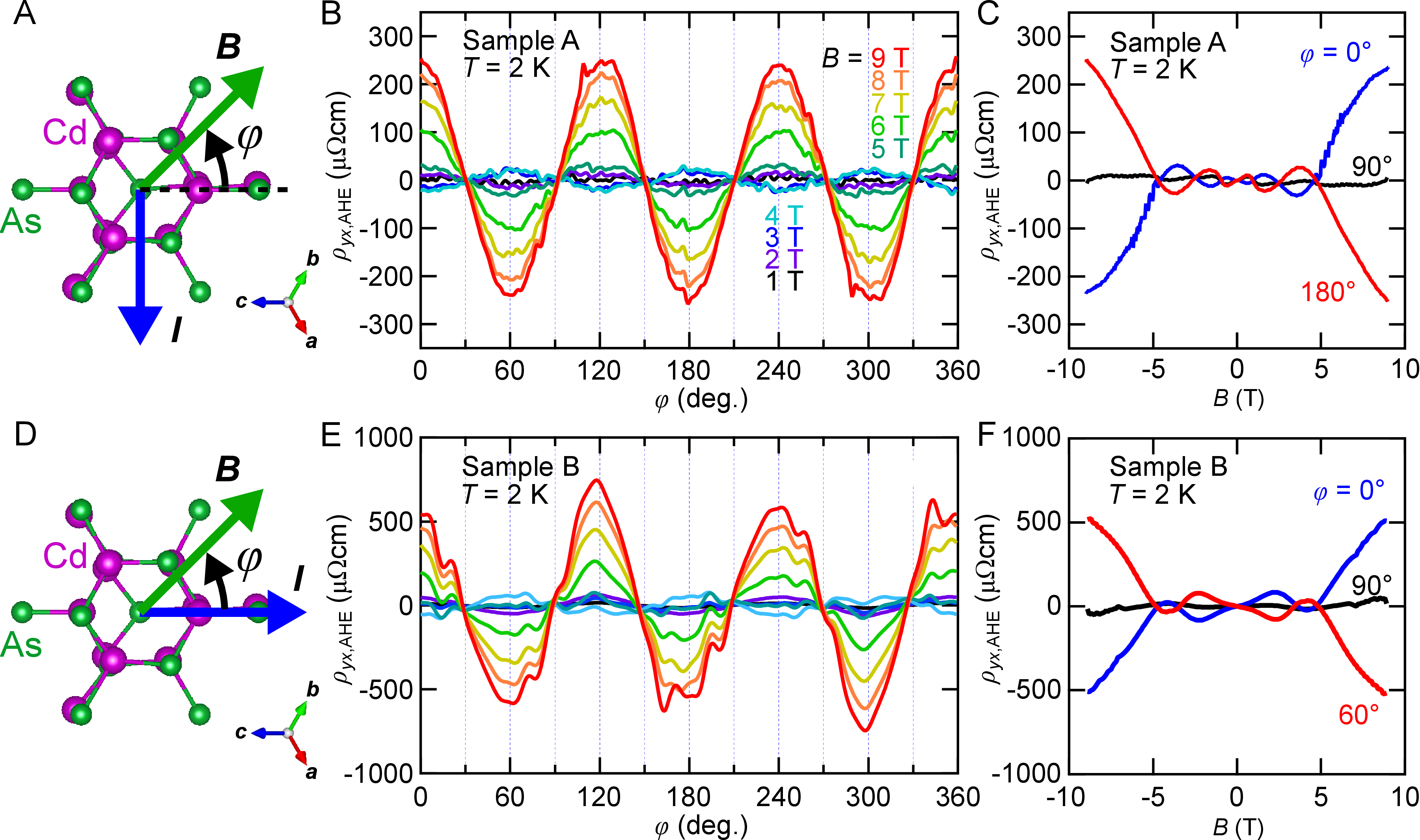}
\caption{In-plane anomalous Hall effect observed in {\CA} films. The azimuthal angle of the in-plane field $\varphi$ is measured from [$11\bar{4}$]. (\textit{A}) Measurement configuration for Sample A with the current applied along the direction of $\varphi = 270^{\circ}$. (\textit{B}) $\varphi$ dependence of in-plane anomalous Hall resistivity {\rhoyxAHE} measured at different fields, and (\textit{C}) field dependence of {\rhoyxAHE} at different $\varphi$ for Sample A. (\textit{D}) Measurement configuration for Sample B with the current applied along the direction of $\varphi = 0^{\circ}$. (\textit{E}) $\varphi$ dependence and (\textit{F}) field dependence of {\rhoyxAHE} for Sample B.
}
\label{fig3}
\end{center}
\end{figure*}

\begin{figure*}
\begin{center}
\clearpage
\newpage
\includegraphics[width=16cm]{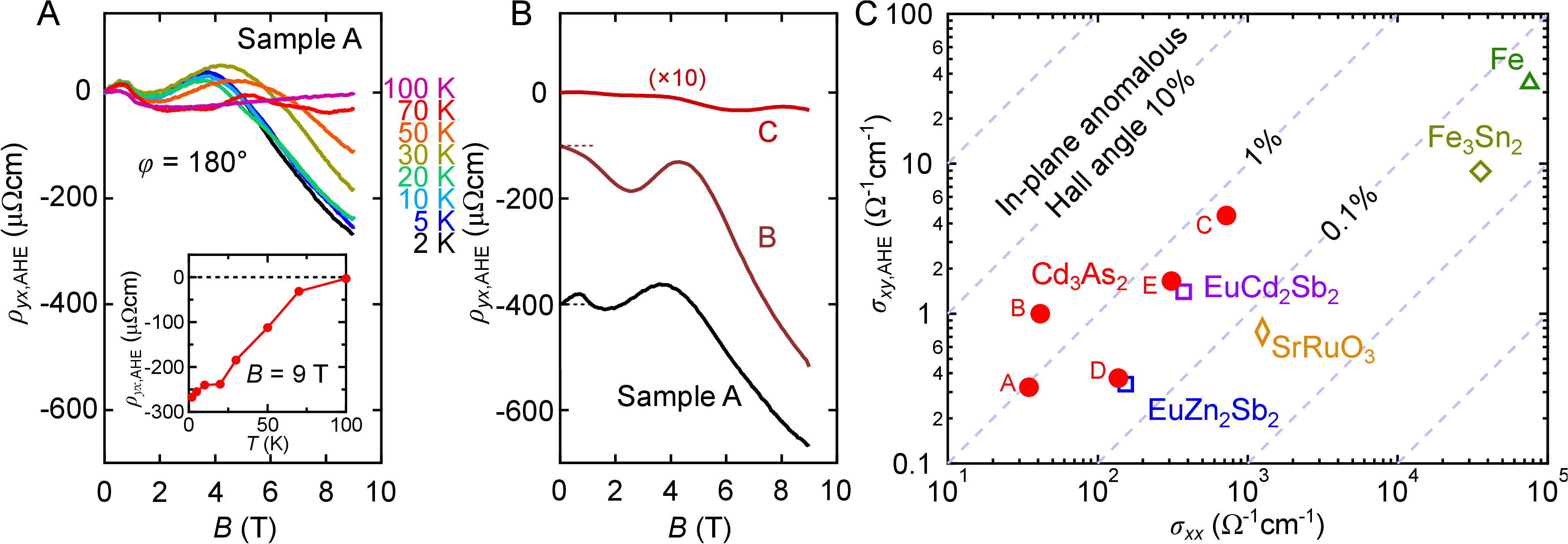}
\caption{
Temperature and carrier density dependence of in-plane AHE in the {\CA} films. (\textit{A}) Temperature dependence of {\rhoyxAHE} measured at $\varphi = 180^{\circ}$ for Sample A. Inset shows the change of the {\rhoyxAHE} value at 9 T which exhibits a monotonic decay with increasing temperature. (\textit{B}) Comparison of the field dependence of {\rhoyxAHE} to the higher-electron-density film (Sample C). Data are shifted for clarity. (\textit{C}) Relation between the in-plane anomalous Hall conductivity {\sigmaxyAHE} and longitudinal conductivity {\sigmaxx} for the {\CA} films, compared to other magnetic systems exhibiting in-plane AHE with three-fold symmetry: {\ECS}\cite{iAHE_ECS}, {\EZS}\cite{iAHE_EZS}, {\FS}\cite{iAHE_FeSn}, Fe\cite{iAHE_Fe}, and {\SRO}\cite{iAHE_SRO}.
}
\label{fig4}
\end{center}
\end{figure*}

%
\end{document}